*Effects of low energy electron irradiation on formation of nitrogen-vacancy centers in single-crystal diamond*

J. Schwartz[1], S. Aloni[2], D. F. Ogletree[2], and T. Schenkel[1]

[1]Accelerator and Fusion Research Division, Lawrence Berkeley National Laboratory, Berkeley CA 94720, USA
[2]The Molecular Foundry, Lawrence Berkeley National Laboratory, Berkeley CA 94720, USA

Exposure to beams of low energy electrons (2 to 30 keV) in a scanning electron microscope locally induces formation of NV-centers without thermal annealing in diamonds that have been implanted with nitrogen ions. We find that non-thermal, electron beam induced NV-formation is about four times less efficient than thermal annealing. But NV-center formation in a consecutive thermal annealing step (800 °C) following exposure to low energy electrons increases by a factor of up to 1.8 compared to thermal annealing alone. These observations point to reconstruction of nitrogen – vacancy complexes induced by electronic excitations from low energy electrons as an NV-center formation mechanism and identify local electronic excitations as a means for spatially controlled room-temperature NV-center formation.

1. **Introduction**

Negatively charged nitrogen-vacancy centers (NV$^-$) in diamond are promising quantum bit candidates [1-3] and sensitive probes for high resolution magnetometry [4, 5]. Emerging applications require techniques for reliable formation and placement of NV-centers with long spin coherence times and high degrees of spectral stability. Nitrogen is a common impurity in diamond. In Ib and IIa type diamonds, nitrogen is present as isolated substitutional centers (P1 centers). The widely accepted model for NV-center formation is the trapping of a vacancy by a substitutional nitrogen atom during thermal annealing [6-10]. Vacancies are created through irradiation with energetic neutrons, electrons or ions [11, 12]. Here, the incident particles must transfer sufficient energy, ~35 eV, to displace carbon atoms from their lattice positions and create vacancies and carbon interstitials [13]. This requires e. g. electrons with energies >150 keV [13]. Following vacancy creation, diamonds are commonly annealed above ~ 600°C where vacancies become mobile [7, 14]. Mobile vacancies are then trapped by substitutional nitrogen, forming NV-centers [7]. The charge state of NV-centers (NV$^-$ or NV$^0$) is sensitive to the local environment as NV$^-$ centers capture their additional electron e. g. from nearby substitutional nitrogen atoms or surface states [10, 15, 16].

For many applications and areas of research it is highly desirable to form NV-centers at precise locations in highly pure diamonds with minimal nitrogen background. Nitrogen has therefore been introduced into diamond by ion implantation [8, 10, 17-19]. Vacancies and carbon interstitials are generated through collisions as incident keV to MeV nitrogen ions stop in the diamond target. Most nitrogen ions come to rest on interstitial positions, but a small fraction may also be incorporated into substitutional lattice sites through replacement collisions [20]. Again, thermal annealing above 600°C is employed to form NV-centers [8, 21]. A high conversion efficiency of implanted nitrogen into NV-centers is desirable because fluctuations in the nitrogen spin bath are a source of decoherence [9, 22]. For NV-center formation in nitrogen implanted diamond it has always been assumed that during thermal annealing nitrogen atoms are first incorporated into substitutional sites in the diamond lattice, followed by the trapping of a mobile vacancy by a substitutional nitrogen atom. In this article we report that exposure to low energy electrons at room temperature also forms NV-centers in nitrogen implanted diamond, without any thermal annealing. Further, we find that exposure to low energy electrons followed by thermal annealing increases NV-formation efficiencies significantly compared to thermal annealing alone.

## 2. Experimental Methods

Electronic grade, low nitrogen background (5<ppb) synthetic CVD diamonds from Element 6 [23] were cleaned by 10 min sonication in acetone and isopropyl alcohol followed by 10 min boiling in Piranha at 120 °C. After cleaning, masked mm size regions of samples were implanted with $10^{12}$ $N^+/cm^2$ at 7.7 keV normal to the (100) direction for a mean depth of 12 nm [19, 24, 25]. Calibrated electron beam exposures were made at grids of single points and in raster-scanned squares using a Zeiss Supra field-emission scanning electron microscope equipped with a pattern generator and fast beam blanker. Incident electron energies ranged from 2 to 30 keV with a beam focused to 3 nm or better at currents of 9 pA to 40 nA as determined by a Faraday cup. Hyper spectral optical images (full spectrum at each pixel, 40 to 100 ms acquisition time) were acquired before and after e-beam exposures with a WITec confocal micro-Raman system equipped with a grating spectrometer and CCD camera using an 0.95 NA air objective and a 532 nm laser delivering 13.6 mW to the sample. The optical spectra showed diamond Raman lines and photoluminescence (PL) from NV-centers in either neutral ($NV^0$) or negative ($NV^-$) charge states. The acquisition software could be used to generate 2-D images from selected spectral regions after spectrometer background subtraction. After nitrogen implantation, e-beam exposure and optical characterization, some samples were annealed in vacuum (base pressure low $10^{-7}$ Torr) at temperatures of 400°C (25 min.) or 800°C (15 min.) and re-imaged. Some samples were also exposed to beams of low energy electrons while held at temperatures of 400, 660 and 800°C.

## 3. Results

In Figure 1, we show PL images of a high purity diamond generated by integrating PL spectra from each image pixel around the $NV^-$ zero-phonon line (ZPL, 635 to 642 nm). The diamond had been implanted with nitrogen ions and 1 $\mu m^2$ areas were then exposed to a 2 keV, 9 pA electron beam at room temperature for exposure times ranging from 0.1 to 100 s. $NV^-$ centers were formed with electron doses as low as 3 pC/$\mu m^2$ (~0.5 mC/$cm^2$) without thermal annealing. No $NV^-$ centers were formed by electron exposure of pristine diamond areas that had not been implanted with nitrogen ions.

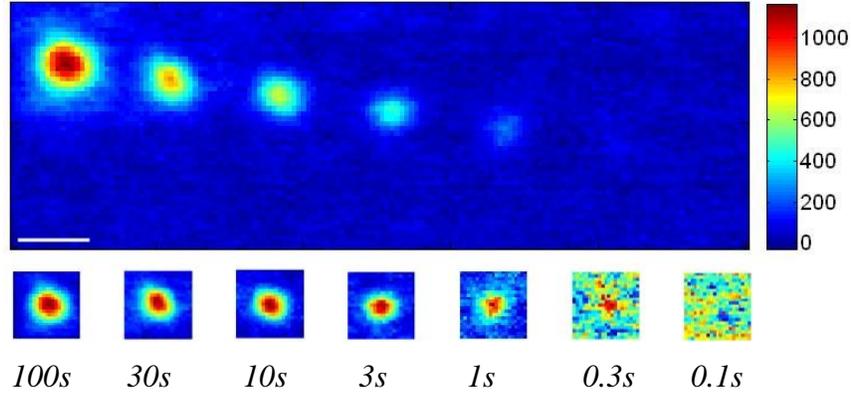

100s  30s  10s  3s  1s  0.3s  0.1s

*Figure 1: Confocal PL image of NV⁻ centers (integrated spectral intensity 635 to 642 nm). The image was recorded following exposure of 1μm squares with a 2 keV, 9 pA electron beam. Insets show locally auto-scaled spot details. The scale bar is 3 μm.*

We then performed a series of electron beam exposures with varying doses and electron beam energies on similar diamond samples to elucidate mechanisms of non-thermal NV⁻ formation. In Figure 2 we show the mean NV⁻ PL intensity within uniformly-irradiated 1 μm² areas as a function of dose for a 2 keV electron beam. Areas exposed to the same dose with beam currents varying by two orders of magnitude show similar brightness. The NV⁻ density saturates at high dose. The intensities for different dose rates fall near one common dose curve, indicating that there was no strong dose-rate effect. Similar data were obtained for exposures with 5 keV electrons (not shown).

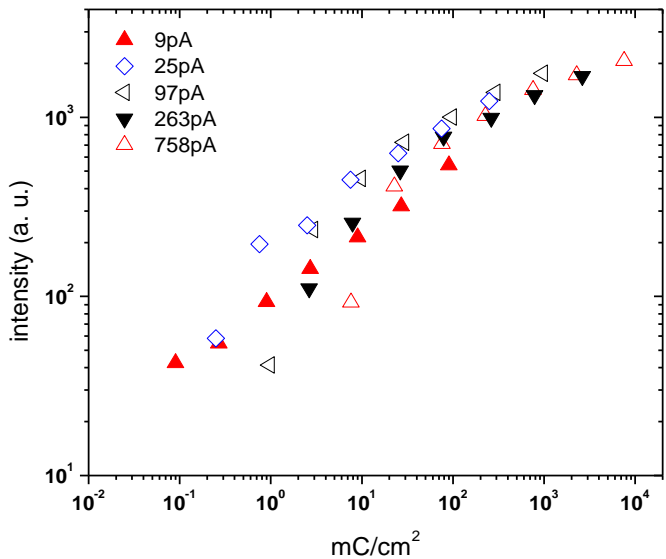

*Figure 2: NV⁻ PL intensity from 1μm squares exposed to 2 keV electrons as a function of accumulated dose (1 mC/cm² =10 pC/μm² =6.2×10¹⁵ e⁻/cm²). The symbols correspond to different beam currents, each data set with exposure times of 0.1s, 0.3s, 1s, 3s, 10s, 30s and 100s.*

In Figure 3 we show the relative NV⁻ PL intensity as a function of incident electron beam energy at one constant dose of 800 mC/cm². A maximum of $1.8 \cdot 10^{-4}$ of the incident electron kinetic energy, $E_{kin}$, can be transferred to carbon atoms in non-relativistic collisions, i. e. ≤0.36 eV for 2 keV electrons, far below the ~35 eV energy transfer threshold to create knock-on vacancies in diamond [13]. The increase and then saturation of the NV⁻ PL intensity as a function of electron beam energy is due to the formation of electron – hole pairs and the density of electronic excitation in the shallow sample volume where implanted nitrogen atoms are present (see discussion below).

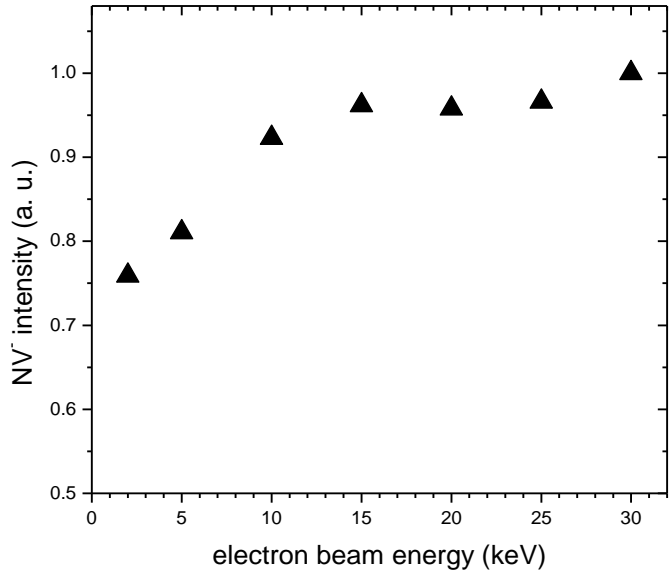

*Fig 3: NV⁻ PL-intensity as a function of electron beam energy. $3 \times 5 \mu m^2$ areas were exposed to 40 nA electron beams with a common dose of 800 mC/cm² each.*

We conducted a series of thermal anneals of nitrogen-implanted diamonds along with electron beam exposures to explore the interplay of electron-beam induced and thermal annealing effects on NV-center formation. Samples were exposed to electron beams before, during or after annealing at temperatures between 400 and 800 °C. Very low intensities of NV's were observed after annealing at 400°C. The largest increase in NV⁻ PL signals was observed for exposure to 10 keV electrons with doses above 1 C/cm² at room temperature, followed by annealing at 800°C for 15 min. Under this condition, the NV⁻ signal is up to 1.8 times higher than for annealing at 800°C only, without exposure to electrons (Figure 4). The NV⁻ PL signal intensity is also ~4 times larger after 800°C annealing than after room-temperature exposure with a saturation dose of 10 keV electrons alone, without thermal annealing. E-beam exposures had no effect on NV-signals following annealing at 800°C. The absolute efficiency for conversion of implanted nitrogen into NV⁰ and NV⁻ centers was not measured here. From results reported by Pezzagna et al. [9] for low energy nitrogen ion implantation we estimate that the absolute conversion efficiency was about 1 to 2%. NV-center charge states are known to be affected by surface conditions [15]. The relative yields we observe for different processing conditions in Figure 4 were similar for $NV^0$ and $NV^-$. The observed increase of relative NV-intensity is encouraging and points to potential further enhancements that might be realized by optimization of electron beam exposures together with thermal annealing and co-implantation [10, 24, 26]. Following formation of NV-centers, their charge states can be controlled by e. g. by surface passivation [15] or electrical gating.

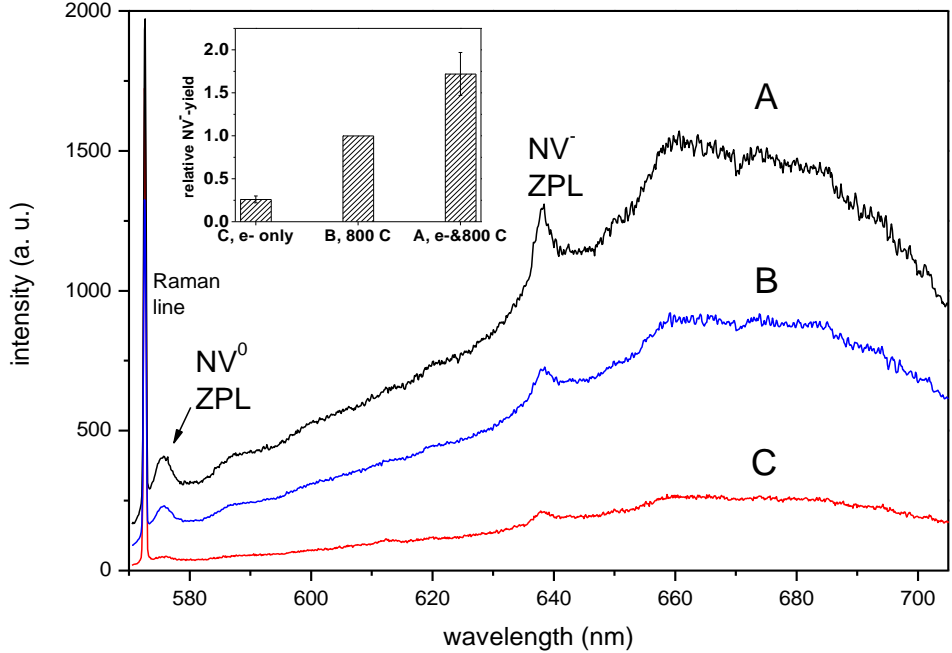

*Figure 4: PL spectra (room temperature, excitation at 532 nm) from a nitrogen ion implanted diamond sample for (A) 10 keV, 65 C/cm$^2$ electron beam exposure followed by 800°C anneal, B: annealing at 800°C alone with no electron beam exposure, C: electron beam exposure only, no annealing. Varying noise levels are due to averaging conditions. The insert summarizes relative NV PL intensities for the different processing conditions. While the PL measurement errors are small, we observed reproducible spatial variations in PL intensities, possibly due to surface effects.*

4. **Discussion**

We now consider mechanisms of low-energy electron assisted NV-center formation. Both direct momentum-transfer to lattice atoms and beam-induced thermal heating can be ruled out. As mentioned above, direct momentum transfer in small impact parameter collisions can only transfer a small fraction of the kinetic energy of incident electrons to carbon atoms in the diamond lattice, ≤5.4 eV for 30 keV electrons, the highest electron beam energy used here. This value is well below the threshold energy for vacancy formation in diamond at room temperature of ~35 eV, which requires electron energies above 150 keV [13].

Electron beam heating can be estimated by assuming that energy is uniformly deposited in a sphere of radius R~30 nm at 2 kV, then $\delta T=(I \cdot V)/(\pi \cdot k \cdot R)$ (eq. 1), where k is the thermal conductivity of diamond (2000 W/m·K). Thus $\delta T \leq 0.01$ K/nA at 2 keV and even less at higher $E_{kin}$ since $R \sim E_{kin}^{1.75}$ [27].

Energetic electrons interacting with solids loose energy primarily through inelastic valence-electron scattering events transferring ~10-20 eV, creating hot electrons and holes. The mean electron-hole pair creation energy in diamond is ~13 eV/e-h pair. Hot carriers thermalize within a few ps to states near the conduction and valence band minima, where they diffuse before recombining by radiative or non-radiative processes. Some carriers are trapped by surface or point defects [28]. For 2 keV electrons, the pair generation rate peaks ~20 nm below the surface at ~$1.8 \cdot 10^{10}$ /nA·nm·s. Higher energy electrons generate more carriers, but they also penetrate more deeply into the lattice, resulting in a lower carrier creation density. Following ion implantation, nitrogen atoms are distributed along a depth profile with a peak concentration of ~$9 \cdot 10^{17}$ cm$^{-3}$ about 12 nm below the surface and a width of approximately 10 nm, slightly broadened by ion channeling [19]. The electron dose required to form NV-centers does not change much with increasing $E_{kin}$ (Figure 3) because higher energy electrons deposit an increasing part of their energy outside of the nitrogen implanted sample volume.

Electron-exposed regions were broadened by more than the PL image resolution for both spot and rectangular exposures (Figure 1). Exposed spots, for example, had a full width at half max (FWHM) of 1.2 µm, while a diffraction-limited PL spot should have a FWHM of ~0.6 µm. The electron beam was focused to a few nm spot and its energy was deposited in a radius of 30 nm or less [21]. The observation of NV-center creation outside of directly-irradiated areas points strongly to a mechanism based on carrier creation and electronic excitation. The observed "blooming" is caused by carriers diffusing out of the directly irradiated area.

NV-center formation requires the motion of point defects within the diamond lattice. Many defects in diamond, including the NV-center, show significant electron-phonon coupling, i. e., the equilibrium atomic geometry changes with excitation or charge state [29, 30]. When defects capture electrons or holes the resulting changes in atomic configuration transfer energy to the lattice. When excited $NV^0$ or $NV^-$ centers relax, up to ~ 0.5 eV may be transferred to lattice vibrations, significantly more than the thermal energy at 800 °C. While diamond is being irradiated by electrons, as in cathodoluminescence spectroscopy, we observed light emission almost exclusively from NV-centers in the neutral charge state. We did not observe surface charging effects during electron beam exposures, likely because our sample mounting clips provided sufficiently effective grounding paths. Strong electron-phonon coupling provides a plausible mechanism for electron-induced reconstruction of meta-stable defect, e. g. comprised of interstitial nitrogen surrounded by several vacancies. Electronic excitations, delivered by energetic light ions [31, 32] or electrons [33], has been shown to aid repair of crystal damage and induce the dissociation of boron-deuterium pairs in diamond.

Long-range point-defect diffusion outside of regions irradiated by high-energy electrons has been reported [34].

Very detailed theoretical and experimental studies have elucidated the nature of many radiation induced defects in diamond, their structure and stability [11, 34-38]. We have found that electron-beam exposure of nitrogen implanted diamond samples forms NV-centers without annealing as well as during and after annealing at 400 and 660 °C, but not after annealing at 800 °C. With the widely used Monte Carlo model SRIM [25] we can estimate that about 50 vacancies are created per implanted nitrogen ion (7.7 keV), along with carbon interstitials. A significant fraction of defects recombine following the cooling of collision cascades but the detailed point defect dynamics, which is at the core of the problem of efficient NV-center formation, is unknown. The implanted N are mostly interstitials although a fraction may occupy substitutional sites due to replacement collisions [20]. It seems unlikely that carrier capture promotes vacancy diffusion, since many studies have used MeV electron irradiation to produce vacancies and in all cases thermal annealing was required to produce NV-centers [6, 11, 35]. Enhanced diffusion of N interstitials in different charge states has been suggested by theoretical studies [39] and recent electron paramagnetic resonance studies have identified nitrogen interstitial related defects [37]. One possible scenario is that a defect complex containing an interstitial nitrogen atom near two vacancies reconstructs upon electronic excitation by low energy electrons such that the interstitial nitrogen recombines with one vacancy and traps another to form a stable $NV^-$ or $NV^0$ center. Low energy electrons also affect the charge states of defect complexes, which might impact NV-formation together with electronic excitation. Nitrogen interstitials may also relax into different configurations, possibly involving vacancies and carbon interstitials and become trapped in defect configurations such that no further motion can be induced by electronic excitation and carrier capture. Annealing at 800 °C dissolves this defect complex and consecutive electron beam exposure has no effect on NV-intensities. It is intriguing to note that the relative NV-center yield for electron beam exposure followed by thermal annealing is larger then the sum effects of electron beam exposure (without annealing) and thermal annealing (without electron exposure). Apparently, exposure to low energy electrons leads to formation of some NV-centers and it also makes it more likely that more NV-centers are formed during an ensuing thermal annealing step. We speculate that the latter might be related to the neutralization of charged defects that were created in the ion implantation process. Trapping of nitrogen atoms in stable or hard to dissolve defect centers is likely limiting achievable NV-center formation efficiencies. Some of these defects are probably not PL active and may [37] or may not be paramagnetic and thus elude experimental access.

The ultimate spatial resolution of electron-beam induced NV-center formation depends on carrier distributions, which are influenced by carrier lifetime and surface recombination rates. The latter are difficult to determine for un-annealed diamond after nitrogen ion implantation. Low energy electron exposure generates carriers in an approximately spherical volume and carrier concentration drops off outside the generation regions as $1/r^2$. Higher energy electron beams (> 20 keV) will penetrate deeper into the sample, and act as a line source of carriers, whose concentration will drop off as $1/r$ outside the exposed area. In electron beam lithography this latter effect is used to achieve a resolution of 10 nm or better [40].

5. Conclusion

In conclusion, we report effects of low energy electrons on formation of NV-centers in nitrogen implanted single crystal diamonds. Exposure to low energy electrons alone leads to formation of NV-centers without any thermal annealing, but with low efficiency. Exposure to low energy electrons, followed by thermal annealing at 800° C enhances NV-center formation by a factor of 1.8 compared to thermal annealing alone. The formation of NV-centers by non-thermal reconstruction of nitrogen – vacancy complexes induced by electronic excitations points to an alternative mechanism for NV-center formation vs. trapping of a vacancy by a substitutional nitrogen atom. Electron beam induced NV-formation is a "direct write" technique that could become useful in applications where thermal annealing is prohibitive or undesirable, e. g. in combination of diamonds with biological systems, delicate waveguides or photonic cavities. The interplay of electronic excitation and thermal annealing points to directions for further process optimization, e. g. together with co-implantation [24, 26] and optimized thermal annealing protocols [10]. With adaptation of single ion implantation techniques to nitrogen ions (e. g. via detection of secondary electrons or the sensing of surface current changes) [18], local e-beam activation can also enable formation of arrays of single NV-centers in diamond with high spatial resolution.

**Acknowledgements**

This work was performed in part at the Molecular Foundry and the National Center for Electron Microscopy, Lawrence Berkeley National Laboratory and was supported by the Office of Science, Office of Basic Energy Sciences, Scientific User Facilities Division, of the U.S. Department of Energy under Contract No. DE-AC02—05CH11231. This work was also supported by the Laboratory Directed Research and Development Program of Lawrence Berkeley National Laboratory under the same contract, and by the DARPA Quest program through a subcontract from UC Santa Barbara.

**Corresponding author contact information:** T_Schenkel@lbl.gov, 510-486-6674